\documentclass[11pt]{article}
\usepackage[utf8]{inputenc}
\usepackage{amsmath,amsfonts,amssymb}
\usepackage{amsthm}
\usepackage{latexsym}
\usepackage[noadjust]{cite}
\usepackage{graphicx}
\usepackage{xcolor}
\usepackage{dirtytalk}
\usepackage{mathtools}
\usepackage[margin=1in]{geometry}
\usepackage{thm-restate}
\usepackage{enumitem}
\usepackage[linesnumbered,ruled]{algorithm2e}
\usepackage[colorlinks=true, allcolors=blue]{hyperref}
\usepackage[nameinlink,capitalise]{cleveref}
\hypersetup{
    citecolor={violet}
}
\pdfoutput=1

\newtheorem{theorem}{Theorem}[section]
\newtheorem{corollary}[theorem]{Corollary}

\theoremstyle{definition}
\newtheorem{definition}[theorem]{Definition}
\newtheorem{remark}[theorem]{Remark}

\newcommand{\R}{\mathbb{R}}

\newcommand{\calG}{\mathcal{G}}

\newcommand{\liftH}{\psi(H)}

\newcommand{\zo}{\{0, 1\}}

\newcommand{\encode}{\mathrm{Encode}}
\newcommand{\decode}{\mathrm{Decode}}

\newcommand{\eps}{\varepsilon}

\newcommand{\ra}{\rightarrow}

\newcommand{\cut}{\text{cut}}
\newcommand{\head}{\text{head}}
\newcommand{\tail}{\text{tail}}

\newif\ifdraft
\drafttrue

\newif\ifnotanon
\notanontrue

\title{Almost-Tight Bounds on Preserving Cuts in Classes of Submodular Hypergraphs}
\ifnotanon{
\author{Sanjeev Khanna\thanks{School of Engineering and Applied Sciences, University of Pennsylvania, Philadelphia, PA. Email: {\tt sanjeev@cis.upenn.edu}. Supported in part by NSF awards CCF-1934876 and CCF-2008305.} \and Aaron (Louie) Putterman\thanks{Supported in part by the Simons Investigator Award of Madhu Sudan and NSF Award CCF 2152413. Supported in part by a Simons Investigator Award of Salil Vadhan. Supported in part by a Hudson River Trading PhD Research Scholarship. Email: \texttt{aputterman@g.harvard.edu}.} \and Madhu Sudan\thanks{School of Engineering and Applied Sciences, Harvard University, Cambridge, Massachusetts, USA. Supported in part by a Simons Investigator Award and NSF Award CCF 2152413. Email: \texttt{madhu@cs.harvard.edu}.}}}
\else
\fi
\date{\today}

\begin{document}

\maketitle

\pagenumbering{gobble}

\begin{abstract}

Recently, a number of variants of the notion of cut-preserving hypergraph sparsification have been studied in the literature. These variants include directed hypergraph sparsification, submodular hypergraph sparsification, general notions of approximation including spectral approximations, and more general notions like sketching that can answer cut queries using more general data structures than just sparsifiers. In this work, we provide reductions between these different variants of hypergraph sparsification and establish new upper and lower bounds on the space complexity of preserving their cuts. Specifically, we show that:
\begin{enumerate}
    \item $(1 \pm \epsilon)$ directed hypergraph spectral (respectively cut) sparsification on $n$ vertices efficiently reduces to  $(1 \pm \epsilon)$ undirected hypergraph spectral (respectively cut) sparsification on $n^2 + 1$ vertices. Using the work of Lee and Jambulapati, Liu, and Sidford (STOC 2023) this gives us directed hypergraph spectral sparsifiers with $O(n^2 \log^2(n) / \epsilon^2)$ hyperedges and directed hypergraph cut sparsifiers with $O(n^2 \log(n)/ \epsilon^2)$ hyperedges by using the work of Chen, Khanna, and Nagda (FOCS 2020), both of which improve upon the work of Oko, Sakaue, and Tanigawa (ICALP 2023).
    \item Any cut sketching scheme which preserves all cuts in any directed hypergraph on $n$ vertices to a $(1 \pm \epsilon)$ factor (for $\epsilon = \frac{1}{2^{O(\sqrt{\log(n)})}}$) must have worst-case bit complexity $n^{3 - o(1)}$. Because directed hypergraphs are a subclass of submodular hypergraphs, this also shows a worst-case sketching lower bound of $n^{3 - o(1)}$ bits for sketching cuts in general submodular hypergraphs. 
    \item $(1 \pm \eps)$ monotone submodular hypergraph cut sparsification on $n$ vertices efficiently reduces to $(1 \pm \eps)$ symmetric submodular hypergraph sparsification on $n+1$ vertices. Using the work of Jambulapati et. al. (FOCS 2023) this gives us monotone submodular hypergraph sparsifiers with $\widetilde{O}(n / \eps^2)$ hyperedges, improving on the $O(n^3 / \eps^2)$ hyperedge bound of Kenneth and Krauthgamer (arxiv 2023).
\end{enumerate}
At a high level,  our results use the same general principle, namely, by showing that cuts in one class of hypergraphs can be simulated by cuts in a simpler class of hypergraphs, we can leverage sparsification results for the simpler class of hypergraphs.

\end{abstract}

\newpage
\pagenumbering{arabic}

\section{Introduction}

Sparsification deals with the following natural question: given a large object, how much can we compress it while still retaining some of its key properties? In the realm of graphs, this has been a well-studied notion spanning decades of research. Starting with the work of Karger \cite{Kar93}, the question of how sparse we can make a graph while still preserving the approximate sizes of every cut has been a central topic of research. Since then, numerous works by many authors have resolved this question (starting with the work of Bencz\'ur and Karger \cite{BK96}) and pushed the boundaries of this research beyond just graph cuts \cite{BSS09, ST11, KK15, CKN20}. 

More rigorously, for a weighted graph $G = (V, E)$ on $n$ vertices, we can define a cut in the graph corresponding to each set $S \subseteq V$. For such a set $S$, we define the vector $\mathbf{1}_S \in \zo^{|V|}$ as the indicator vector of whether the $i$th vertex is in $S$. Using this vector, we say that $\cut_G(S) = \sum_{(u, v) \in E} w_{(u, v)} ({\mathbf{1}_S}_u - {\mathbf{1}_S}_v)^2$, i.e., the weight of the edges crossing between $S$ and $V-S$. A cut-sparsifier asks for a reweighted subset of edges $\hat{E} \subseteq E$ such that in the graph $G = (V, \hat{E})$, with the corresponding new weights $\hat{w}$, for every $S \subseteq V$
\[
(1 - \eps) \cut_G(S) \leq \cut_{\hat{G}}(S) \leq (1 + \eps) \cut_G(S).
\]
The seminal work of \cite{BK96} was the first to show the existence of such sparsifiers $\hat{G}$ for any graph $G$ such that $|\hat{E}| = \widetilde{O}(n / \eps^2)$. Subsequent work in the \emph{spectral} regime asked whether such sparsifiers still exist when we consider real-valued vectors as opposed to cut-vectors. In this setting, we define a Laplacian $L_G$ for our graph $G$. We say that for $x \in \R^{|V|}$
\[
x^T L_G x =  \sum_{(u, v) \in E} w_{(u, v)} (x_u - x_v)^2.
\]
 The goal in this regime instead becomes finding a reweighted subgraph $\hat{G}$ such that for every $x \in \R^{|V|}$,
\[
(1 - \eps) x^T L_{\hat{G}} x \leq  x^T L_{G} x \leq (1 + \eps) x^T L_{\hat{G}} x.
\]
Work by Batson, Spielman, and Srivastava, and Spielman and Teng \cite{BSS09, ST11} settled the size complexity of spectral sparsifiers for ordinary graphs by showing the existence of such sparsifiers of size $O(n / \eps^2)$.

Recently, starting with the work of Kogan and Krauthgamer \cite{KK15}, a natural extension to the study of graph sparsification has been the study of sparsifying \emph{hypergraphs}. In this setting, one is given a hypergraph $H = (V, E)$, and asked to preserve to a $(1 \pm \eps)$ factor the weight of all hyperedges crossing a particular cut. A cut is given by a bichromatic coloring of the vertices and a hyperedge is considered cut if it is not monochromatic.
Work by Chen, Khanna, and Nagda \cite{CKN20} was the first to completely characterize the cut-sparsifiability of hypergraphs, which showed that there exist $(1 \pm \eps)$-cut-sparsifiers for any hypergraph on $n$ vertices of size  $O(n \log (n) / \eps^2)$. As in the graph setting, where the natural next step from cut-sparsifiers was spectral-sparsifiers, Soma and Yoshida \cite{SY19} later introduced this notion of spectral \emph{hypergraph} sparsification. More explicitly, the \say{energy function} (also called the Laplacian) of an undirected hypergraph $H = (V, E)$ is as follows:
\[
x^T L_H x = \cut_H(x) = \sum_{e \in E} w_e \max_{u,v \in e} (x_u - x_v)^2.
\]
A $(1 \pm \eps)$-spectral sparsifier for an undirected hypergraph then corresponds to a reweighted subhypergraph of $H$, denoted by $\hat{H}$ such that for any $x \in \R^{|V|}$, 
\[
(1 - \eps)x^T L_H x \leq x^T L_{\hat{H}} x \leq (1 + \eps)x^T L_H x.
\]
This question of whether one could preserve the Laplacian of undirected hypergraphs with only a near-linear number of hyperedges was then resolved by Kapralov et. al. \cite{KKTY21b}, Jambulapati, Liu, and Sidford \cite{JLS22}, and Lee \cite{Lee23} in the affirmative. 

More recently however, work has sought to generalize hypergraph sparsification even further. Indeed, given a hypergraph $H = (V, E)$, instead of viewing edge-cuts in the traditional way (i.e., for a bichromatic coloring of the vertices counting how many hyperedges are not one color), a more general \emph{splitting} function is assigned to each hyperedge $e \subseteq V$. This splitting function is a set function $g_e: 2^{e} \ra \R^{\geq 0}$. One natural extension to the case of ordinary hypergraphs that has received particular attention is the case in which these splitting functions $g_e$ are also required to be submodular \cite{KK23, KZ23} (though there has also been work on the regime where these functions are not submodular, for instance with parity functions in \cite{KPS24}). For such a submodular hypergraph $H = (V, E)$, the value on any cut $S \subset V$ is 
\[
\cut_H(S) = \sum_{e \in E} g_e(S \cap e).
\]

Recall that a function $g: 2^V \ra \R^{\geq 0}$ is said to be \emph{submodular} if it has the property of diminishing returns. That is, for any $S \subset T \subset V$, and any element $x \in V, x \notin T$,
\[
g(S \cup \{ x \}) - g(S) \geq g(T \cup \{ x \}) - g(T).
\]

Under this definition, one type of submodular hypergraph is a \emph{directed hypergraph}. In a directed hypergraph, one can view each directed hyperedge instead as a tuple $(e_{\tail}, e_{\head})$ of subsets of $V$. The cut function of a directed hyperedge $e$ on cut $S$ is $1$ if and only if an element from $S$ is in $e_{\tail}$ and an element from $V - S$ is in $e_{\head}$. More explicitly, for a directed hypergraph $G = (V, E, w)$ on $n$ vertices, and a vector $x \in \R^n$, we can define the Laplacian for $G$ as
	\[
	x^T L_G x = \sum_{e \in E} \max_{u \in L(e), v \in R(e)} (x_u - x_v)_+^2.
	\]
	In this context, $(x_u - x_v)_+ = \max((x_u - x_v), 0)$, and directed hypergraph cuts are simply the restriction of the vector $x$ to be in $\zo^{|V|}$ (seen as the indicator vector for a set $S \subseteq V$). A non-zero contribution from a hyperedge occurs only if a tail vertex of the hyperedge has a larger value than a head vertex of the hyperedge. 

One can check that in the cut regime (i.e. $x \in \zo^n$), each directed hyperedge cut yields a submodular function $g_e: 2^{e_{\head} \cup e_{\tail}} \ra \R^{\geq 0}$. In what follows, we describe our contributions to various problems in this area.

\subsection{Improved Bounds for Directed Hypergraph Sparsification}

In the graph case, it is known that directed graph cut-sparsifiers for graphs with $n$ vertices can require as many as $\Omega(n^2)$ edges to preserve cuts to a $(1 \pm \eps)$ factor. In this sense, directed graph cut-sparsification is a trivial task, as any graph has at most $O(n^2)$ edges to begin with. Contrary to this however, directed \emph{hypergraph} sparsification is non-trivial. While the same $\Omega(n^2)$ lower bound exists, a directed hypergraph can have as many as $4^n$ directed hyperedges to start with, so a sparsifier with $O(n^2)$ directed hyperedges is a vast improvement. This observation has led to a rich line of research studying the feasibility of sparsifying directed hypergraphs. The first work on this front was the work of \cite{SY19} which showed the existence of directed hypergraph sparsifiers with $O(n^3/\eps^2)$ directed hyperedges and gave a polynomial time algorithm for computing them. Later work by \cite{KKTY21b} presented a proof of sparsifiers with $\widetilde{O}(n r / \eps^2)$ (where $r$ is the maximum size of any hyperedge) hyperedges for undirected hypergraph spectral sparsification, and with $\widetilde{O}(n^2 r^3 / \eps^2)$ directed hyperedges for directed hypergraph spectral sparsification by tuning their algorithm and performing a different analysis. In particular, this improved upon the result of \cite{SY19} in the regime where $r$ is constant. Note that as with graphs, spectral sparsification is a \emph{stronger} notion than cut sparsification, so in particular, these proofs imply the existence of cut-sparsifiers of the same complexity.

Ultimately however, the complexity of directed spectral hypergraph sparsification was nearly settled by the work of Oko, Sakaue, and Tanigawa \cite{OST23}, who showed $(1 \pm \eps)$ spectral-sparsifiers with $O(n^2 \log^3(n / \eps) / \eps^2)$ directed hyperedges exist for directed hypergraphs on $n$ vertices. 

Continuing this line of research, we show that fundamentally, the task of directed hypergraph sparsification can be reduced in a black-box manner to undirected hypergraph spectral sparsification.

More specifically, we show there is a lifting from a directed hypergraph on $n$ vertices to an undirected hypergraph on $n^2 + 1$ vertices such that the Laplacian of every individual hyperedge is simultaneously preserved. That is, we show the following theorem:

\begin{theorem}\label{thm:directedReduction}
	For $H = (V, E)$ an a directed hypergraph on $n$ vertices, one can compute an undirected hypergraph $\psi(H)$ on $n^2+1$ vertices in time $O(mr^2)$ (where $m$ is the number of hyperedges in $H$, and $r$ is the maximum size of any hyperedge in $H$), such that for any $x \in \R^n$, one can also compute $\vartheta(x) \in \R^{n^2 + 1}$ in time $O(n^2)$ such that 
 \[
 x^T L_H x = \vartheta(x)^T L_{\psi(H)} \vartheta(x).
 \]
 
 Moreover, for any hyperedge $e \in H$, there is a single corresponding hyperedge $\psi(e)$ in $\psi(H)$ such that
	\[
	x^T L_e x = \vartheta(x)^T L_{\psi(e)} \vartheta(x).
	\]
	The size of $\psi(e)$ is at most $|e|^2$. Further, for $x \in \zo^n$, i.e. corresponding to a cut, $\vartheta(x)$ will be in $ \zo^{n^2+1}$, i.e. also corresponding to a cut.
\end{theorem}

We can then use the existing state of the art literature of undirected spectral hypergraph sparsification \cite{JLS22,Lee23} to conclude the existence of directed spectral hypergraph sparsifiers with only $O(n^2 \log(n) \log(r) / \eps^2)$ hyperedges which can be found in time $\widetilde{O}(m r^2)$, where $m$ is the original number of hyperedges and $r$ is the maximum size of any hyperedge. Note that this bound on the size of sparsifiers improves on the result of \cite{OST23}, and in particular, makes the dependence on $\eps$ exactly $O(1 / \eps^2)$, which now matches the literature for undirected sparsification. That is, we show the following:

\begin{theorem}
	For any directed hypergraph $H = (V, E)$ on $n$ vertices, and any $0 < \eps < 1$ there exists a weighted sub-hypergraph $\hat{H}$ such that for all $x \in \R^n$:
	\[
	(1  - \eps) x^T L_H x \leq x^T L_{\hat{H}}x \leq (1  + \eps) x^T L_H x,
	\]
	and $\hat{H}$ only has $O(n^2 \log(n) \log(r) / \eps^2)$ hyperedges, where $r$ is the maximum size of any hyperedge of $H$. 
\end{theorem}

As an additional benefit, because the reduction of \cref{thm:directedReduction} preserves cut vectors, we can also invoke the result of \cite{CKN20} to conclude the existence of directed hypergraph \emph{cut}-sparsifiers with $O(n^2 \log(n) / \eps^2)$ hyperedges.

\subsection{Lower Bounds for Sketching Cuts in Directed Hypergraphs}

We next focus on the bit complexity of creating cut-sparsifiers for directed hypergraphs. This is done in hopes of answering an open question from \cite{KK23} regarding the bit-complexity of arbitrary sketching schemes for submodular hypergraphs. In prior work \cite{OST23,KK23}, a lower bound of size $\Omega(n^3)$ (ignoring $\eps$) was established for the bit complexity of any directed hypergraph cut-\emph{sparsifier}. However, lower bounds for sparsifiers explicitly take advantage of the sparsifier structure by starting with known examples of sparsifiers that require $\Omega(n^2)$ hyperedges, and then padding these hyperedges with random vertices in their tail such that the bit complexity of each hyperedge becomes $\Omega(n)$. One can trivially show that this padding does not change the requirement of preserving $\Omega(n^2)$ hyperedges.
Because sparsifiers are limited to storing only hyperedges that were originally present, this then forces a bit complexity lower bound of $\Omega(n^3)$. However, this same technique is not amenable to a sketching lower bound as the padding procedure only adds complexity to each hyperedge, and not necessarily to the cut function as a whole. Thus, the difficulty is in showing that the cut function itself requires a large description size, regardless of how we choose to represent it. This marks a fundamental difference.

Addressing this, we show the following theorem:

\begin{theorem}
    Any $(1 \pm \eps) $ cut-sketching scheme for directed hypergraphs on $n$ vertices must have worst-case space $\frac{n^3}{2^{O(\sqrt{\log(n)})}}$ bits (for $\eps = \frac{1}{2^{O(\sqrt{\log(n)})}}$).
\end{theorem}

At a high level, our proof takes advantage of a result of Kapralov et. al.  \cite{KKTY21a}. In this work, the authors show that there exists a family of undirected hypergraphs on $n$ vertices, each with at most $n$ hyperedges, such that any sketching scheme which can sketch cuts in any of the hypergraphs in this family to an additive error of $\eps n$ (for $\eps = \frac{1}{2^{O(\sqrt{\log(n)})}}$) must have worst-case size at least $\frac{n^2}{2^{O(\sqrt{\log(n)})}}$. We show that by using a specific construction of a directed hypergraph, along with a specific reconstruction procedure, we can actually store an additive cut-approximation to $n$ distinct undirected hypergraphs in a single cut-sketch of a directed hypergraph. That is, we show the following theorem:

\begin{theorem}
    For any undirected hypergraphs $H_1, \dots H_n$, each on vertex set $V$, with $|V| = n$, there exists a directed hypergraph $G$ on $2n$ vertices, such that given a $(1 \pm \eps)$ cut-sketch for $G$, for any of the undirected hypergraphs $H_i = (V, E_i)$, one can recover $\cut_{H_i}(S)$ to within additive error $3\eps |E_i|$.
\end{theorem}

Now, by sampling these undirected hypergraphs $H_1, \dots H_n$ from a specific family of hypergraphs, we can argue that simultaneously preserving the cut-values in all of these hypergraphs (even to an additive error) requires a data structure of size $\frac{n^2}{2^{O(\sqrt{\log(n)})}} \cdot n = \frac{n^3}{2^{O(\sqrt{\log(n)})}}$. In particular, by the previous reduction, any general scheme for sketching directed hypergraphs or submodular hypergraphs would be such a scheme, and therefore must have worst-case size at least $\Omega(n^{3-o(1)})$ (for $\eps = \frac{1}{2^{O(\sqrt{\log(n)})}}$). 

Prior to our work, there was no known super-quadratic (in $n$) lower bound on the sketching complexity of cuts in directed hypergraphs. In conjunction with our positive results on the sparsifiability of directed hypergraphs, this shows that directed hypergraph sparsification is almost-optimal even among all possible sketches for preserving cut values. That is, from the previous section, we know that directed hypergraph sparsifiers approximately preserve the sizes of all cuts in a directed hypergraph to a factor $(1 \pm \eps)$ using $\widetilde{O}(n^3 / \eps^2)$ bits. In conjunction with our lower bound, we can conclude that this is almost the best possible (among \emph{any} sketching scheme) in the regime where $\eps = \frac{1}{2^{O(\sqrt{\log(n)})}}$. Thus, we show that for approximately storing cuts in directed hypergraphs using as few bits as possible, using a sparsifier is almost optimal. We view this as an important contribution of our work.

\subsection{Cut Sparsifiers for Monotone Submodular Hypergraphs}

Finally, we show that one can simulate cuts in monotone submodular hypergraphs with cuts in symmetric submodular hypergraphs. Recall that a set function is monotone if $f(S \cup \{ t \}) \geq f(S)$, and we say that a submodular hypergraph is monotone if \emph{every} splitting function is also monotone. This model of hypergraphs was specifically studied in the work of \cite{KK23}, where their sparsifiers ultimately achieved a complexity of $O(n^3 / \eps^2)$ hyperedges. In particular, monotone submodular functions capture a wide variety of natural and common functions such as matroid rank and entropy of random variables.

With respect to this, we show the following theorem:

\begin{theorem}
    Suppose $f: 2^V \ra \R^{\geq 0}$ is a monotone, submodular function. Then, $f': 2^{V \cup \{ * \}} \ra \R^{\geq 0}$ defined as $\forall S \subseteq V$
    \[
    f'(S) = f(S) = f'(V - S \cup \{ * \})
    \]
    is submodular and symmetric.
\end{theorem}

Next, we show that given an arbitrary monotone, submodular hypergraph on $n$ vertices, we can lift this to a symmetric submodular hypergraph on $n+1$ vertices, where the single extra vertex is the $\{*\}$ vertex from the preceding theorem. Next, for each individual splitting function $g_e: 2^e \ra \R^+$ in the monotone, submodular hypergraph, we replace $g_e$ with $g_e'$, again using the preceding theorem. 

Note that for each monotone submodular function, we \emph{re-use} the same $\{*\}$ vertex. Thus, the increase in the size of the vertex set is only $1$. Finally, we can then invoke a result from \cite{JLLS23}, which states that for any submodular hypergraph $H$ where each splitting function is symmetric, one can calculate a sparsifier for $H$ with only $\widetilde{O}(n / \eps^2)$ hyperedges.

We then get the following:

\begin{theorem}
    Let $H = (V, E)$ be a hypergraph, such that $\forall e \in E$, the corresponding splitting function $g_e: 2^e \ra \R^{\geq 0}$ is submodular and monotone. Then there exists a $(1 \pm \eps)$ cut-sparsifier for $H$ retaining only $\widetilde{O}(n/ \eps^2)$ hyperedges.
\end{theorem}

Prior to this work, the best known upper bound for the size complexity (in hyperedges) for $(1 \pm \eps)$-sparsifying any monotone submodular hypergraph was $O(n^3 / \eps^2)$ hyperedges, proved in the work of \cite{KK23}. Our result essentially improves this to the best possible, where we now only have a near-linear dependence on the size of the vertex set. We view it as an interesting open question if one can extend our proof method used here to general submodular functions (although this case will necessarily require a blow-up of at least quadratic size).

\subsection{Overview}

At a high level, all of our results use the same general principle, namely, by showing that cuts in one class of hypergraphs can be simulated by cuts in a simpler class of hypergraphs, we can leverage sparsification results for the simpler class of hypergraphs. This leads to our proofs being quite simple despite the fact that the results improve upon the state-of-the-art knowledge in hypergraph sparsification.

In \cref{sec:prelim} we introduce formal definitions and other preliminaries. In \cref{sec:dirtoundir} we present the algorithms for sparsifying directed hypergraphs by reducing to undirected hypergraph sparsification. Next, in \cref{sec:lowerbounds}, we show how to simultaneously simulate cuts in many different undirected graphs thereby leading to new lower bounds for the worst case size of sketching cuts in directed hypergraphs. Finally, in \cref{sec:monotone}, we show how to sparsify arbitrary monotone, submodular hypergraphs to near-optimal size.

\section{Preliminaries}\label{sec:prelim}

First, we introduce the definitions of undirected and directed hypergraphs.

\begin{definition}
	An \textbf{undirected hypergraph} $G = (V, E)$ is a collection of vertices $V$, with associated hyperedges $e \in E$, where $e \subseteq V$ can be of arbitrary size. 
\end{definition}

\begin{definition}
	A \textbf{directed hypergraph} $H = (V, E)$ is a collection of vertices $V$ along with directed hyperedges $e \in E$. Each directed hyperedge is of the form $e = (e_{\text{tail}}, e_{\text{head}})$, where $e_{\head}, e_{\tail} \subseteq V$. We will use $L(e) =e_{\text{tail}}, R(e) =  e_{\text{head}}$. Note that $e_{\head}, e_{\tail}$ are not necessarily disjoint. 
\end{definition}

Next, we introduce the definition of spectral sparsifiers for both undirected and directed hypergraphs. 

\begin{definition}
	For an undirected hypergraph $G = (V, E, w)$ on $n$ vertices, and a vector $x \in \R^n$, the \textbf{quadratic form of the Laplacian of $G$} is 
	\[
	x^T L_G x = \sum_{e \in E} \max_{u, v \in e} (x_u - x_v)^2.
	\]
\end{definition}

\begin{definition}
	For a directed hypergraph $G = (V, E, w)$ on $n$ vertices, and a vector $x \in \R^n$, the \textbf{directed quadratic form of the Laplacian of $G$} is
	\[
	x^T L_G x = \sum_{e \in E} \max_{u \in L(e), v \in R(e)} (x_u - x_v)_+^2.
	\]
	In this context, $(x_u - x_v)_+ = \max((x_u - x_v), 0)$. A non-zero contribution from a hyperedge occurs only if a head vertex of the hyperedge has a larger value than a tail vertex of the hyperedge. Note that the head set and tail set of a directed hyperedge are not necessarily disjoint.
\end{definition}

\begin{definition}
	For a (directed or undirected) hypergraph $G = (V, E)$ on $n$ vertices, a \textbf{$(1 \pm \eps)$-spectral sparsifier} for $G$ is a weighted (directed or undirected) sub-hypergraph $H$ such that for every $x \in \R^n$, 
	\[
	(1 - \eps) x^T L_G x \leq x^T L_H x \leq (1 + \eps) x^T L_G x.
	\]
	
	Further, we require that the hyperedges of $H$ are a subset of the hyperedges of $G$. 
\end{definition}

\begin{remark}
    For all the above definitions, if a reweighted sub-hypergraph $H$ of $G$ preserves the quadratic form for vectors $x \in \zo^n$ to $(1 \pm \eps)$ multiplicative error, we say that $H$ is a \textbf{cut-sparsifiers}. Note that all spectral sparsifiers are cut-sparsifiers, while the converse is not necessarily true. 

    We also refer to cut-sizes in hypergraphs. A cut is specified by a set $S \subseteq V$, and we say the size of the cut $S$ in $G$ (denoted $|\cut_G(S)|)$ is $ (\mathbf{1}_S)^T L_G (\mathbf{1}_S)^T$, where $\mathbf{1}_S$ is the indicator vector in $\zo^n$ for the set $S$. Combinatorially, this refers to the weight of the hyperedges that are \say{leaving} the set $S$.
\end{remark}

Next we define submodular functions and submodular hypergraphs.

\begin{definition}
    A function $g: 2^V \ra \R^{\geq 0}$ is said to be submodular if for any $S \subset T \subset V$, and any $x \in V - T$,
    \[
    g(S \cup \{ x \}) - g(S) \geq g(T \cup \{ x \}) - g(T).
    \]
\end{definition}

Using this, we can define a submodular hypergraph.

\begin{definition}
    A submodular hypergraph $H = (V, E)$ is a set of $n$ vertices along with a set of hyperedges $E$. For each hyperedge $e \in E$, there is a corresponding submodular splitting function $g_e: 2^e \ra \R^{\geq 0}$. For any subset $S \subseteq V$, the corresponding cut of the submodular hypergraph is 
    \[
    \cut_H(S) = \sum_{e \in E} g_e(S \cap e).
    \]
\end{definition}

\begin{definition}
    We say that a data structure $G$ is a $(1 \pm \eps)$-cut \emph{sketch} of a submodular hypergraph $H = (V, E)$, if for any $S \subseteq V$ one can deterministically recover $\cut_H(S)$ to within a $(1 \pm \eps)$ factor using only the data structure $G$, and the set $S$.
\end{definition}

We will use the following result from \cite{JLLS23} regarding the sparsifiability of symmetric, submodular hypergraphs. Note that a submodular function $f: 2^V \ra \R^+$ is said to be symmetric if $\forall S \subseteq V, f(S) = f(V-S)$.

\begin{theorem}\label{thm:sparseSymmetric}[Corollary 1.2 of \cite{JLLS23}]
    For any symmetric submodular hypergraph $H$ on $n$ vertices, there is a $(1 \pm \eps)$-sparsifier for $H$ with $\widetilde{O}(n / \eps^2)$ hyperedges.
\end{theorem}

\section{Directed to Undirected Hypergraph Sparsification}\label{sec:dirtoundir}

In this section, we will show that any algorithm that produces an undirected spectral hypergraph sparsifier with $f(n, r)$ hyperedges (for a vertex set of size $n$, and maximum hyperedge size $r$), can be used in a {\em black-box manner} to create a spectral sparsifier with $f(n^2 + 1, r^2)$ hyperedges for any $n$-vertex directed hypergraph.

To this end, we first have to define the \say{lifting} operation from a directed hypergraph on $n$ vertices to an undirected hypergraph on $n^2 + 1$ vertices.

\begin{definition}
	For a directed hypergraph $H = (V, E)$ on $n$ vertices, let $\liftH$ be an undirected hypergraph on $n^2 + 1$ vertices defined as follows. For the first $n^2$ vertices of $\liftH$, associate these vertices with tuples of vertices from $H$, that is, each of these vertices is associated with an element from the set $V \times V$. The final vertex in $\liftH$ will be a special vertex we denote by $*$. Now, for each hyperedge $e \in E$ of $H$, define a corresponding hyperedge $\varphi(e)$ in $\liftH$ as follows: let the vertices in $L(e)$ be $u_1, \dots u_{\ell}$, and let the vertices in $R(e)$ be $v_1, \dots v_r$. Let $\varphi(e)$ contain 
 \[ L(e) \times R(e) \cup \{ *\} = \{(u_1, v_1), (u_1, v_2), \dots (u_1, v_r), (u_2, v_1), \dots (u_2, v_r), \dots (u_3, v_1), \dots (u_{\ell}, v_{r}), * \}.\]
	
	Note that this transformation is invertible. If we are given an undirected hyperedge of the form $\varphi(e) =  L(e) \times R(e) \cup \{ *\}$, we can invert this transformation to recover the directed hyperedge $e = (L(e), R(e))$. Additionally, note that this transformation and its inverse are efficiently computable (running in time $O(r^2)$, where $r$ is the size of the undirected hyperedge). 
\end{definition}

Next, we define the lifting of a test vector.

\begin{definition}
	For a vector $x \in \R^n$, we define the lifting of $x$ denoted as $\vartheta(x)$. $\vartheta(x)$ is in $\R^{n^2 + 1}$, and in particular, for the first $n^2$ entries, we associate these with the set $[n] \times [n]$. We say that $(\vartheta(x))_{u, v} = \max(x_u - x_v, 0)$. For the final entry, which we associate with the special vertex $*$ in the lifted $H$, we let $\vartheta(x)_{*} = 0$. 
\end{definition}

Note again that $\vartheta(x)$ is efficiently computable in time $O(n^2)$ where $n$ is the dimension of $x$.

\begin{theorem}
	Let $H = (V, E)$ be a directed hypergraph on $n$ vertices. Then, for any $x \in \R^n$,
	\[
	x^T L_H x = \vartheta(x)^T L_{\liftH} \vartheta(x).
	\]
\end{theorem}

\begin{proof}
	It suffices to show that for a single hyperedge $e \in E$,
	\[
	\max_{u \in L(e), v \in R(e)} (x_u - x_v)_+^2 = \max_{(y, z) \in \varphi(e)} (\vartheta(x)_y - \vartheta(x)_z)^2.
	\]
	The reason this suffices is that there is one $\varphi(e)$ for each corresponding hyperedge $e \in E$. So, we are in effect showing that every term in the sum of the quadratic form of the Laplacians is the same.
	
	To see why this equality is true,  let some $\widehat{u} \in L(e), \widehat{v} \in R(e)$ be the maximizers for the expression on the left. Then, note that the corresponding entry $\vartheta(x)_{\widehat{u}, \widehat{v}}$ is exactly $(x_{\widehat{u}} - x_{\widehat{v}})_+$. Now, because $\widehat{u} \in L(e)$ and $\widehat{v} \in R(e)$, it follows that $(\widehat{u}, \widehat{v}) \in \varphi(e)$. Because the special vertex $* \in \varphi(e)$, it follows that in the above expression
	\[
	\max_{(y, z) \in \varphi(e)} (\vartheta(x)_y - \vartheta(x)_z)^2 \geq (\vartheta(x)_{(\widehat{u}, \widehat{v})} - \vartheta(x)_{*})^2 = (x_{\widehat{u}} - x_{\widehat{v}})_+^2 = \max_{u \in L(e), v \in R(e)} (x_u - x_v)_+^2.
	\] 
	
	Now, we will show the opposite direction. Indeed, suppose that some elements $\widehat{y}, \widehat{z}$ are maximizers for $\max_{(y, z) \in \varphi(e)} (\vartheta(x)_y - \vartheta(x)_z)^2$. Note that by construction, every entry in $\vartheta(x)$ is $\geq 0$. This means that without loss of generality, we can assume that $\widehat{z} = *$ (the special vertex), as this vertex attains the smallest possible value $0$. This means that the maximizing value of the expression is exactly $\vartheta(x)_{\widehat{y}}^2$, where $\widehat{y}$ is one of the first $n^2$ vertices in $\liftH$. So, let us write $\widehat{y} = (\widehat{a}, \widehat{b})$, where $\widehat{a}, \widehat{b}$ are both vertices in $G$. By construction, because $\widehat{y} \in \widehat{e}$, it follows that $\widehat{a} \in L(e)$, and $\widehat{b} \in R(e)$. As such it follows that 
	\[
	\max_{u \in L(e), v \in R(e)} (x_u - x_v)_+^2 \geq (x_{\widehat{a}} - x_{\widehat{b}})_+^2 = \vartheta(x)_{\widehat{y}}^2 =  \max_{(y, z) \in \widehat{e}} (\vartheta(x)_y - \vartheta(x)_z)^2.
	\]
	
	Thus, it follows that 
	\[
	\max_{u \in L(e), v \in R(e)} (x_u - x_v)_+^2 = \max_{(y, z) \in \varphi(e)} (\vartheta(x)_y - \vartheta(x)_z)^2,
	\]
	as claimed.
\end{proof}

\begin{corollary}
	Let $H$ be a directed hypergraph on $n$ vertices. Suppose that $\widehat{\liftH}$ is a $(1 \pm \eps)$ undirected hypergraph spectral sparsifier to $\liftH$. Then, it follows that the unlifted graph $\widehat{H}$ which is calculated by applying $\varphi^{-1}$ to each hyperedge in $\widehat{\liftH}$, is a $(1 \pm \eps)$ directed hypergraph spectral sparsifier to $H$.
\end{corollary}

\begin{proof}
	Indeed, suppose $H, \liftH, \widehat{H}, \widehat{\liftH}$ are as specified above, and let $x \in \R^n$. It follows that 
	\[
	(1 - \eps ) x^T L_H x = (1 - \eps) \vartheta(x)^T L_{\liftH} \vartheta(x) \leq \vartheta(x)^T L_{\widehat{\liftH}} \vartheta(x) = x^T L_{\widehat{H}} x \]
 \[= \vartheta(x)^T L_{\widehat{\liftH}} \vartheta(x) \leq (1+\eps) \vartheta(x)^T L_{\liftH} \vartheta(x) = (1 + \eps ) x^T L_H x.
	\]
	
	To conclude, this implies that for $\widehat{H}, H$ as above,
	\[
	(1 - \eps ) x^T L_H x \leq x^T L_{\widehat{H}} x \leq (1 + \eps ) x^T L_H x.
	\]
\end{proof}

\begin{theorem}
	For a directed hypergraph $H$ on $n$ vertices, one can find a directed hypergraph spectral sparsifier $\widehat{H}$ of $H$, with $O(n^2 \log(n) \log(r) / \eps^2)$ hyperedges in time $\widetilde{O}(m r^2)$, where $m$ is the number of directed hyperedges in $H$ and $r$ is the maximum size of a hyperedge in $H$.
\end{theorem}

\begin{proof}
	If the number of hyperedges in $H$ is less than $n^2$, simply return $H$. Otherwise, lift $H$ to $\liftH$, and spectrally sparsify $\liftH$ using \cite{JLS22}. This will result in a $(1 \pm \eps)$ spectral sparsifier $\widehat{\liftH}$ to $\liftH$, with at most $O(n^2 \log(n^2) \log(r^2) / \eps^2)$ hyperedges, as we desire. Here, we have used that the maximum rank of a hyperedge in $\widehat{G}$ is at most the squared rank of a hyperedge in $G$. Further, the running time of this algorithm is $\widetilde{O}(m r^2)$, as the number of hyperedges in $\widehat{G}$ is the same as the number of hyperedges in $G$, and the rank, again, is at most $r^2$. Now, we can unlift $\widehat{\liftH}$ to $\widehat{H}$ by applying $\varphi^{-1}$ to each hyperedge, and use the previous corollary to conclude our theorem.
\end{proof}

\begin{remark}
    Note that if we restrict our original vector $x$ to be in $\zo^n$, it follows that $\vartheta(x) \in \zo^{n^2 + 1}$. By repeating the exact same steps above, this means that we can use the same reduction from above to get directed hypergraph cut sparsifiers, by only using algorithms from undirected hypergraph cut sparsifiers. 
\end{remark}

\begin{corollary}\label{cor:dirCutSparse}
    For a directed hypergraph $H$ on $n$ vertices, one can find a directed hypergraph cut sparsifier $\widehat{H}$ of $H$, with $O(n^2 \log(n) / \eps^2)$ hyperedges in time $\widetilde{O}(mr^2 / \eps^2)$, where $m$ is the number of directed hyperedges in $H$.
\end{corollary}

\begin{proof}
    Simply perform the reduction from above, and invoke the algorithm for undirected hypergraph cut-sparsification from \cite{Qua23}.
\end{proof}

\newcommand{\tildeH}{\widetilde{H}}
\newcommand{\tildeG}{\widetilde{G}}
\newcommand{\newSplit}{\hat{g'_e}}

\section{Space Lower-bounds for Sketching Cuts in Directed Hypergraphs}\label{sec:lowerbounds}

In this section, we will establish an $\Omega(n^{3-o(1)})$ lower-bound for worst-case sketching of the cuts in a directed hypergraph on $n$ vertices to a $(1 \pm \eps)$ factor for $\eps$ being $\frac{1}{2^{O(\sqrt{\log(n)})}}$. As mentioned in the introduction, this improves upon a result of \cite{KK23} who showed a lower bound of size $\Omega(n^3)$ for the bit complexity of any \emph{sparsifier}. However, their lower bound explicitly takes advantage of the sparsifier structure by starting with known examples of sparsifiers that require $\Omega(n^2)$ hyperedges, and then padding these hyperedges with random vertices in their tail such that the bit complexity is $\Omega(n)$. One can trivially show that this padding does not change the requirement of preserving $\Omega(n^2)$ hyperedges.
Because sparsifiers are limited to storing only hyperedges that were originally present, this then forces a bit complexity lower bound of $\Omega(n^3)$. However, this same technique is not amenable to a sketching lower bound as the padding procedure only adds complexity to each hyperedge, and not necessarily to the cut function as a whole.

To overcome this, we take advantage of a result of \cite{KKTY21a} who showed that, in general, any $(1 \pm \eps)$ cut-sketching scheme for undirected hypergraphs on $n$ vertices, with $\eps = \frac{1}{2^{O(\sqrt{\log(n)})}}$ must have worst case bit complexity $\frac{n^2}{2^{O(\sqrt{\log(n)})}}$. This result uses encodings of Rusza-Szemer\'edi graphs into undirected hypergraphs, along with a reconstruction argument to show that general $(1 \pm \eps)$ cut-sketching schemes in undirected hypergraphs give very non-trivial string compression schemes. Then, by invoking known results on size lower bounds for string compression schemes, they are able to conclude worst-case lower bounds of $\frac{n^2}{2^{O(\sqrt{\log(n)})}}$ for the bit complexity of sketching cuts in undirected hypergraphs. To this end, we first reintroduce their notion of a string compression scheme:

\begin{definition}\cite{DN03}
    Let $\ell, k$ be positive integers, and let $\eps, g > 0$. We say that a pair of functions $\encode: \zo^{\ell} \ra \zo^k$ and $\decode: \zo^k \times 2^{[\ell]} \ra \mathbb{N}$ is an $(\ell, k, \eps, g)$ string compression scheme (SCS) if there exists a set of strings $\calG \subseteq \zo^{\ell}$ such that:
    \begin{enumerate}
        \item $|\calG| \geq g \cdot 2^{\ell}$.
        \item For every string $s \in \calG$, and every query $q \in 2^{[\ell]}$, 
        \[
        \left | \decode(\encode(s), q) - |s \cap q| \right | \leq \eps \ell / 2.
        \]
    \end{enumerate}
\end{definition}

The work of \cite{KKTY21a} takes advantage of the following theorem, which is proved in \cite{DN03}:

\begin{theorem}\cite{DN03}\label{thm:dn03}
    Suppose $(\encode, \decode)$ is an $(\ell, k, \eps, g)$-SCS, where $\eps \leq 1/10$. Then,
    \[
    k \geq \frac{\log (g) + 3 \ell / 50}{\log 2} - 1.
    \]
\end{theorem}

Qualitatively, \cite{KKTY21a} shows that for a specific family of \emph{undirected} hypergraphs with $n$ vertices, for some $\eps = \frac{1}{2^{O(\sqrt{\log(n)})}}$ any $(1 \pm \eps)$ cut-sketching scheme for these hypergraphs using $\leq k$ bits implicitly gives an $\left ( \frac{n^2}{2^{O(\sqrt{\log(n)})}}, k, 1/10, 1/2 \right)$-SCS. Thus, by invoking the previous theorem, these sparsifiers must have bit complexity $\frac{n^2}{2^{O(\sqrt{\log(n)})}}$. However, their proof actually provides a stronger result than stated. Although the sparsifiers they use give $(1 \pm \eps)$ multiplicative approximations to cut-sizes, their argument makes uses of an \emph{additive} error bound of $\eps \cdot (\text{\# of hyperedges})$. We take advantage of this in our method by showing that a $(1 \pm \eps)$ cut-sketch for a directed hypergraph can be used to retrieve cut sizes in $n$ distinct undirected hypergraphs with only additive error $\eps$ (with respect to each of these undirected hypergraphs). We first state the result of \cite{KKTY21a} more succinctly, and then describe our construction in more detail. 

\begin{theorem}\cite{KKTY21a}\label{thm:kktySCS}
    For any $n$, and some $\ell = \frac{n^2}{2^{O(\sqrt{\log(n)})}}$, for at least $2^{\ell}/2$ strings $s \in \zo^{\ell}$, there exists an undirected hypergraph $H_s = (V, E_s)$ on $n$ vertices, with $\leq n$ hyperedges, such that any data structure which can approximate cuts in $H_s$ to within additive error $|E_s|/2^{O(\sqrt{\log(n)})}$ can for any query $q \subseteq [\ell]$, answer the subset sum $|q \cap s|$ to within additive error $\ell / 20$.
\end{theorem}

Now, we will prove our theorem regarding capability of directed hypergraphs to simulate cuts in undirected hypergraphs with only additive error.

\begin{theorem}\label{thm:DirHypEncoding}
    Given any undirected hypergraphs $H_1, \dots H_n$, each on vertex set $V$, with $|V| = n$, there exists a directed hypergraph $G$ on $2n$ vertices, such that given a $(1 \pm \eps)$ cut-sketch for $G$, for any of the undirected hypergraphs $H_i = (V, E_i)$ and any set $S \subseteq V$, one can recover $|\cut_{H_i}(S)|$ to within additive error $3\eps |E_i|$.
\end{theorem}

\begin{proof}

As stated, each of the undirected hypergraphs $H_1, \dots H_n$ are on a vertex set of size $n$, which we denote by $V$. We also create a vertex set $W$ of size $n$, which we associate with $w_1, \dots w_n$. Now, we create the directed hypergraph $G$, which lives on the vertex set $V \cup W$ as follows: for each undirected hypergraph $H_i$ for $i = 1, \dots n$, and for each undirected hyperedge $e$ in $H_i$, we add the corresponding directed hyperedge $(e, w_i)$. That is, the head of the directed hyperedge has the vertices from $V$ corresponding to $e$, and the tail of the directed hyperedge has only vertex $w_i$. 

Clearly, $G$ has $2n$ vertices, so now it suffices to argue that for any $H_i = (V, E_i)$, and for any cut $S \subseteq V$, we can recover $\cut_{H_i}(S)$ within additive error $\eps |E_i|$. Indeed, let any such $H_i$ be given, and let $S \subseteq V$ be given as well. Then, suppose we have a $(1 \pm \eps)$ cut-sketch for $G$, which we denote by $\tildeG$. Let us consider the query to $\tildeG$ with the set $S \cup W - \{w_i \}$. A directed hyperedge $e \in G$ is crossing this cut if and only if $e_{\head} \cap (S \cup W - \{w_i \}) \neq \emptyset$ and $e_{\tail} \cap ( (V \cup W) - (S \cup W - \{w_i \})) \neq \emptyset$. In particular, note that by construction, $e_{\head}$ is a subset of $V$ and $e_{\tail}$ is a subset of $W$. This means that a directed hyperedge $e$ is crossing the cut if and only if $e_{\head} \cap S \neq \emptyset$ and $e_{\tail} \cap \{w_i \} \neq \emptyset$. The only directed hyperedges in $G$ which satisfy this second condition are exactly those directed hyperedges in $G$ which correspond to $H_i$. By construction, this means that the number of directed hyperedges crossing this cut $S \cup W - \{w_i \}$ in $G$ is exactly the number of undirected hyperedges $e \in E_i$ such that $e \cap S \neq \emptyset$. Thus this query to $\tildeG$ returns a $(1 \pm \eps)$ approximation to $|\{e \in E_i | S \cap e \neq \emptyset \}|$. Note that the actual size of the cut $S$ in $H_i$ is $|\{e \in E_i | S \cap e \neq \emptyset \land S \cap e \neq e \}|$.

However, note that by symmetry, we can also query $\tildeG$ with $(V-S) \cup (W - \{w_i \})$. By symmetry, this query to $\tildeG$ returns a $(1 \pm \eps)$ approximation to $|\{e \in E_i | (V-S) \cap e \neq \emptyset \}|$, which is the same as $|\{e \in E_i | S \cap e \neq e \}|$. Lastly, we can query $\tildeG$ with $V \cup (W - \{w_i \})$. This query to $\tildeG$ returns a $(1 \pm \eps)$ approximation to $|\{e \in E_i | V \cap e \neq \emptyset \}|$, which is exactly $|\{e \in E_i\}|$.

\newcommand{\satisfies}{\text{ satisfies }}
Now, we operate by the principle of inclusion-exclusion (PIE). Let $A$ be the event that a hyperedge $e \in E_i$ satisfies $e \cap S \neq \emptyset$, and let $B$ be the event that $e$ satisfies $e \cap S \neq e$. By PIE,
\begin{align*}
    |\{e \in E_i | e \satisfies A \land \satisfies B \}| =& |\{e \in E_i | e \satisfies A\}| + |\{e \in E_i | e \satisfies B \}| \\
    & - |\{e \in E_i | e \satisfies A \vee \satisfies B \}|.
\end{align*}

Note that this final expression is trivially satisfied, i.e. $|\{e \in E_i | e \satisfies A \vee \satisfies B \}| = |\{e \in E_i \}|$ as a hyperedge cannot simultaneously have an empty and a non-trivial intersection. Thus, we get that
\begin{align*}
    \cut_{H_i}(S) = & |\{e \in E_i | e \satisfies A \land \satisfies B \}| \\
    =& |\{e \in E_i | e \satisfies A\}| + |\{e \in E_i | e \satisfies B \}|
     - |\{e \in E_i\}|.
\end{align*}

Now, note that our query to $\tildeG$ with the set $S \cup W - \{w_i \}$ gave us a $(1 \pm \eps)$ approximation to $|\{e \in E_i | e \satisfies A\}|$, our query with $(V-S) \cup W - \{w_i \}$ gave us a $(1 \pm \eps)$ approximation to $|\{e \in E_i | e \satisfies B\}|$, and our query with $V \cup W - \{w_i \}$ gave us a $(1 \pm \eps)$ approximation to $|\{e \in E_i \}|$. Because each of these has additive error at most $\eps |E_i|$ (as the error from $\tildeG$ is a multiplicative guarantee), in total, the expression 
\[
\cut_{\tildeG}(S \cup W - \{w_i \}) + \cut_{\tildeG}((V-S) \cup W - \{w_i \}) - \cut_{\tildeG}(V \cup W - \{w_i \})
\]
gives us a $(3\eps |E_i|)$-additive approximation to $\cut_{H_i}(S)$, as we desire.
    
\end{proof}

Now, we will show how we can use the above construction to argue a lower bound of size $\frac{n^3}{2^{O(\sqrt{\log(n)})}}$ on the bit complexity of directed hypergraph cut-sketching. We will do this by showing that we can use a directed hypergraph cut-sketch of size $k$ to create a $(\ell, k, 1/10, 2^{-n})$-SCS, for $\ell = \Omega \left (\frac{n^3}{2^{O(\sqrt{\log(n)})}} \right )$.

\begin{theorem}\label{thm:sketchtoscs}
    A general unweighted directed hypergraph $(1 \pm \frac{1}{2^{O(\sqrt{\log(n)})}})$ cut-sketching scheme on $n$ vertices with maximum sketch size of $k$ bits yields an $(n \cdot \ell, k, 1/10, 2^{-n})$-SCS for $\ell = \frac{n^2}{2^{O(\sqrt{\log(n)})}}$.
\end{theorem}

\begin{proof}
    First, we will define the set $\calG$ of size $\frac{2^{n \cdot \ell}}{2^n}$. Indeed, from \cref{thm:kktySCS}, let $L$ be the strings of length $\ell$ which are able to be compressed and still allow for estimating subset sum queries. Now, let $\calG = L \circ L \circ L \circ \dots \circ L$ ($n$ times), where the $S_1 \circ S_2$ operation takes every string in $S_1$ and prepends it to every string in $S_2$ (resulting in a new set of size $|S_1| \cdot |S_2|$). Note that this means that strings in $\calG$ will be of length $n \cdot \frac{n^2}{2^{O(\sqrt{\log(n)})}}= \frac{n^3}{2^{O(\sqrt{\log(n)})}}$. Further, $\calG$ will be of size $(1/2)^n \cdot 2^{n \cdot \ell}$.

    Now we describe our string compression scheme. Indeed, for any string $s \in \calG$, decompose $s$ into $s_1, \dots s_n$ such that each $s_i \in L$. Now, because each $s_i \in L$, we know there exists a corresponding undirected hypergraph $H_{s_i} = (V, E_{s_i})$ on $n$ vertices such that preserving cuts in $H_{s_i}$ to within additive error $|E_{s_i}|/2^{O(\sqrt{\log(n)})}$ allows us to answer subset sum queries in $H_{s_i}$ to within additive error $\ell/20$. Now let $G$ be the directed hypergraph on $2n$ vertices, built with hypergraphs $H_{s_1}, H_{s_2}, \dots H_{s_n}$ as guaranteed by \cref{thm:DirHypEncoding}. It follows that $G$ is an unweighted directed hypergraph on $2n$ vertices. 

    Now, suppose there exists a general, unweighted, directed hypergraph cut-sketching scheme on $n$ vertices with maximum sketch size of $k$ bits which preserves cuts to a $(1 \pm \frac{1}{2^{O(\sqrt{\log(n)})}})$ multiplicative factor. Then, we can invoke such a scheme on the directed hypergraph $G$ as specified by \cref{thm:DirHypEncoding} to conclude that such a scheme allows us to recover $\cut_{H_{s_i}}(S)$ for any $S \subseteq V$ to within additive error $|E_{s_i}/2^{O(\sqrt{\log(n)})}|$. As a result, this means that for any $s_i$, and any query to $s_i$, denoted by $q_i \in [\ell]$, we can recover $|q_i \cap s_i|$ to within additive error $\ell/20$.

    Finally, suppose we are given any subset query $q \subseteq [n \cdot \ell]$. We want to show that we can compute the size of $|s \cap q|$ (i.e. the sum of the bits of $s$ on the positions indicated by $q$) to within additive error $\frac{n \ell}{20}$. For convenience, we view $q$ as a bit string of length $n \cdot \ell$, where a bit is $1$ if and only if the corresponding element of $[n \cdot \ell]$ was in the subset. Then, we break $q$ into $q_1, \dots q_{n}$ such that each $q_i$ is of length $\ell$. Now, we use the aforementioned sketch to compute $|s_i \cap q_i|$ to within additive error $\ell / 20$ for every $i$. Adding these together, we get an estimate to $|s \cap q|$ with additive error at most $n \ell / 20$. Thus, a general directed hypergraph cut-sketching scheme of size $k$ bits to multiplicative error $(1 \pm \frac{1}{2^{O(\sqrt{\log(n)})}})$ yields a $(n \cdot \ell, k, 1/10, 2^{-n})$-SCS.
\end{proof}

\begin{theorem}\label{thm:lowerBoundforDirHyp}
    Any cut-sketching scheme for directed hypergraphs on $2n$ vertices which preserves cuts to a $(1 \pm \eps)$ factor, for $\eps = \frac{1}{2^{O(\sqrt{\log(n)})}}$ must have worst case bit complexity $\frac{n^3}{2^{O(\sqrt{\log(n)})}}$.
\end{theorem}

\begin{proof}
    Indeed, by the preceding theorem (\cref{thm:sketchtoscs}), any such scheme for $\eps = \frac{1}{2^{O(\sqrt{\log(n)})}}$, with bit complexity $k$ implies a $(n \cdot \ell, k, 1/10, 2^{-n})$-SCS, for $\ell = \frac{n^2}{2^{O(\sqrt{\log(n)})}}$. By \cref{thm:dn03} \cite{DN03}, this means that 
    \[
    k \geq \frac{\log(2^{-n}) + 3 n \cdot \ell}{\log 2} -1 \geq \frac{n^3}{2^{O(\sqrt{\log(n)})}}.
    \]
\end{proof}

\begin{corollary}\label{cor:lowerBoundforSubmodHyp}
    Any cut-sketching scheme for submodular hypergraphs on $2n$ vertices which preserves cuts to a $(1 \pm \eps)$ factor, for $\eps = \frac{1}{2^{O(\sqrt{\log(n)})}}$ must have bit complexity $\frac{n^3}{2^{O(\sqrt{\log(n)})}}$.
\end{corollary}

\begin{proof}
    This follows simply by noting that directed hypergraphs are a subclass of submodular hypergraphs, so in particular the lower bound from \cref{thm:lowerBoundforDirHyp} must extend to this case.
\end{proof}

\section{Monotone Hypergraph Sparsifiers}\label{sec:monotone}

In this section, we show how to reduce sparsifying monotone submodular hypergraphs to sparsifying symmetric submodular hypergraphs. At this point, we then invoke the result of \cite{JLLS23} to conclude. First, we detail the reduction:

\begin{theorem}\label{thm:monotoneSymm}
    Suppose $f: 2^V \ra \R^{\geq 0}$ is a monotone, submodular function. Then, $f': 2^{V \cup \{ * \}} \ra \R^{\geq 0}$ defined as $\forall S \subseteq V$
    \[
    f'(S) = f(S) = f'(V - S \cup \{ * \})
    \]
    is submodular and symmetric.
\end{theorem}

\begin{proof}
    First, the symmetry of $f'$ is easy to see. Indeed, for any set $S \subseteq V \cup \{ * \}$, it follows that $f'(S) = f'(V \cup \{ * \} - S)$. So, all that remains to be shown is that $f'$ is submodular. To do this, we will show that $f'$ has decreasing marginals. So consider any $T \subset U \subset V \cup \{ * \}$. We will show that for any $x \notin U$ that 
    \[
    f'(T \cup \{ x \}) - f'(T) \geq f'(U \cup \{ x \}) - f'(U).
    \]
    We will do this by cases.

\begin{enumerate}
    \item Suppose that $x = *$. Then, it must be the case that $x \notin U, T$. So, $f'(U) = f(U), f'(T) = f(T)$. Because $T \subset U$, it must also therefore be the case that $f'(T) \leq f'(U)$ (by the monotonicity of $f$). Next, we note that because $x = *$, $f'(T \cup \{ x \}) = f(V - T),  f'(U \cup \{ x \}) = f(V - U)$. Because $T \subset U$ and $f$ is monotone, it must be the case that $f'(T \cup \{ x \}) = f(V - T) \geq f'(U \cup \{ x \}) = f(V - U)$. Putting this together, we get that it must be the case that 
    \[
    f'(T \cup \{ x \}) - f'(T) \geq f'(U \cup \{ x \}) - f'(U),
    \]
    as we desire.
    \item Suppose that $x \neq *$, and that neither $U, T$ contain $*$. Then the submodularity of $f'$ follows by the submodularity of $f$.
    \item Suppose that $x \neq *$, and that both $U, T$ contain $*$. Then, let $\hat{U}, \hat{T}$ be $U - \{ * \}, T - \{*\}$ respectively. It follows that $\hat{T} \subset \hat{U}$. Further, $f'(T \cup \{ x \}) = f(V - (\hat{T} \cup \{ x \})), f'(U \cup \{ x \}) = f(V - (\hat{U} \cup \{ x \}))$, and likewise $f'(T ) = f(V - \hat{T}), f'(U) = f(V - \hat{U})$. It follows that 
    \begin{align*}
    f'(T \cup \{ x \}) - f'(T) &= f(V - (\hat{T} \cup \{ x \})) - f(V - \hat{T}) \\
    & = f(V - (\hat{T} \cup \{ x \})) - f(V - (\hat{T} \cup \{ x \}) \cup \{ x\})\\
    &\geq f(V - (\hat{U} \cup \{ x \})) - f(V - (\hat{U} \cup \{ x \}) \cup \{ x\}) \\
    &= f'(U \cup \{x\}) - f'(U).
    \end{align*}
    The inequality in the middle holds because $V - (\hat{U} \cup \{ x \}) \subset V - (\hat{T} \cup \{ x \})$. Thus, the marginal gain from adding $x$ to $ V- (\hat{U} \cup \{ x \})$ is larger than the marginal gain from adding $x$ to $V - (\hat{T} \cup \{ x \})$ by the submodularity of $f$.
    \item Suppose that $x \neq *$, but that $* \notin T, * \in U$. Then, by the monotonicity of $f$, $f'(T \cup \{ x\}) - f'(T) = f(T \cup \{ x\}) - f(T) \geq 0$. Likewise, 
    \[
    f'(U \cup \{ x\}) - f'(U) = f(V - (\hat{U} \cup \{x\})) - f(V - \hat{U}) \leq 0,
    \]
    again using the monotonicity of $f$. Therefore, it must be the case that 
    \[
    f'(T \cup \{ x\}) - f'(T) \geq f'(U \cup \{x\}) - f'(U),
    \]
    as we desire.
\end{enumerate}
\end{proof}

Next, we show how to use this reduction to create sparsifiers.

\begin{corollary}
    Let $H = (V, E)$ be a hypergraph, such that $\forall e \in E$, the corresponding splitting function $g_e: 2^e \ra \R^{\geq 0}$ is submodular and monotone. Then there exists a $(1 \pm \eps)$ cut-sparsifier for $H$ with $\widetilde{O}(|V| / \eps^2)$ hyperedges.
\end{corollary}

\begin{proof}
We first define the lifting of a monotone, submodular hypergraph into a symmetric submodular hypergraph.

\begin{definition}
    Let $H = (V, E)$ be a monotone submodular hypergraph. Then, define $H'$ to be the corresponding hypergraph defined on vertex set $V \cup \{ * \}$, where for each edge $e \in E$, we replace it with a hyperedge $e' = e \cup \{ * \}$, and replace the function $g_e$ with the symmetric, submodular splitting function $g_e': 2^{e'} \ra \R^{\geq 0}$ defined in accordance with \cref{thm:monotoneSymm}.
\end{definition}
    
    Now, we construct this hypergraph $H'$. Because each $g_e'$ is symmetric and submodular, we can invoke \cref{thm:sparseSymmetric} to conclude the existence of a hypergraph $\hat{H'}$ such that $\forall S \subseteq V \cup \{ * \}$
    \[
    (1 - \eps) \cut_{H'}(S) \leq \cut_{\hat{H'}}(S) \leq (1 + \eps) \cut_{H'}(S),
    \]
    and $\hat{H'}$ only has $\widetilde{O}(|V| / \eps^2)$ hyperedges remaining.
    
    It follows that because $\forall S \subseteq V$, $g_e'(S) = g_e(S)$, the corresponding hyperedges chosen to create a $(1 \pm \eps)$ cut-sparsifier for $H'$ also create a $(1 \pm \eps)$ cut-sparsifier for $H$. That is, if we create the hypergraph $\hat{H}$ by replacing $e' \in \hat{H'}$ with $e \in H$ (but keeping the same corresponding weights that $\hat{H'}$ assigns), it will be the case that $\forall S \subseteq V$
    \[
    (1 - \eps) \cut_{H'}(S) = (1 - \eps) \cut_H(S) \leq \cut_{\hat{H'}}(S) = \cut_{\hat{H}}(S) \leq (1 + \eps) \cut_{H'}(S) = (1 + \eps)\cut_H(S).
    \]

    Thus, $\hat{H}$ will be a $(1 \pm \eps)$-sparsifier for $H$, and $\hat{H}$ will only keep $\widetilde{O}(|V| / \eps^2)$ hyperedges.
\end{proof}

\bibliographystyle{alpha}
\bibliography{ref}

\end{document}